\documentstyle[12pt]{article}

\newcommand {\vs}[1]  { \vspace*{#1 cm} }

\newcounter{eq}
\newcounter{sc}

\newcommand {\MPL}  {Mod. Phys. Lett.}
\newcommand {\IJMP}  {Int. J. Mod. Phys.}
\newcommand {\NP}   {Nucl. Phys.}

\newcommand {\PL}   {Phys. Lett.}
\newcommand {\PR}   {Phys. Rev.}

\newcommand {\AP}   {Ann. of Phys.}

\def\overleftrightarrow#1{\vbox{\ialign{##\crcr
 $\leftrightarrow$\crcr\noalign{\kern-1pt\nointerlineskip}
 $\hfil\displaystyle{#1}\hfil$\crcr}}}

\newlength{\minitwocolumn}
\begin{document}

\begin{flushright}
DPUR/TH/7\\
January, 2008\\
\end{flushright}
\vspace{30pt}
\pagestyle{empty}
\baselineskip15pt

\begin{center}
{\large\bf Classical Solutions of Ghost Condensation Models

 \vskip 1mm
}

\vspace{20mm}

Masahiro Maeno
          \footnote{
           E-mail address:\ maeno@sci.u-ryukyu.ac.jp
                  }
and Ichiro Oda
          \footnote{
           E-mail address:\ ioda@phys.u-ryukyu.ac.jp
                  }

\vspace{10mm}
          Department of Physics, Faculty of Science, University of the 
           Ryukyus,\\
           Nishihara, Okinawa 903-0213, JAPAN \\

\end{center}


\vspace{20mm}
\begin{abstract}
Motivated by ideas obtained from both ghost condensation and gravitational Higgs mechanism,
we attempt to find classical solutions in the unitary gauge in general ghost condensation
models. It is shown that depending on the form of scalar fields in an action,
there are three kinds of exact solutions, which are (anti-) de Sitter space-time,
polynomially expanding universes and flat Minkowski space-time. We briefly comment
on gravitational Higgs mechanism in these models where we have massive gravitons of 
5 degrees of freedom and 1 unitary scalar field (Nambu-Goldstone boson) after spontaneous
symmetry breakdown of general coordinate reparametrization invariance. The models at
hand are free from the problem associated with the non-unitary propagating mode.

\vspace{15mm}

\end{abstract}

\newpage
\pagestyle{plain}
\pagenumbering{arabic}


\rm
\section{Introduction}

It is known that general relativity successfully describes the gravitational 
phenomena ranging from tiny scales $10^{-4}$ m to huge ones $10^{12}$ m. 
On the other hand, we doubt the validity of general relativity at the
Planck scale owing to appearance of space-time singularity, for which 
quantum effects are expected to play a dominant role. Then, from the 
perspective of duality of length scales it seems to be natural to ask whether 
general relativity might also break down at cosmological distances. 
Actually, general relativity cannot explain some of cosmological phenomena 
such as the accelerating expansion of our universe and gravitational lensing 
without additional inputs like still mysterious dark matter and dark energy etc. 
If these cosmological phenomena were purely of gravitational origin, 
we would have to modify general relativity in the infrared region to some extent.

One interesting approach toward such the direction is to consider the spontaneous 
symmetry breakdown (SSB) of general coordinate reparametrization 
invariance \cite{Percacci}-\cite{'t Hooft}.
The first motivation behind this approach is that SSB might play an important role 
in applying string theory approach to quantum chromodynamics (QCD) in 
future \cite{'t Hooft}. For instance, if we wish to apply a bosonic string theory
to the gluonic sector in QCD, massless fields such as spin 2 gravitons in string theory, 
must become massive or be removed somehow by some ingenious dynamical mechanism 
since such the fields do not exist in QCD.
Note that this motivation is relevant to the modification of general relativity
in the ultraviolet region.

The second motivation comes from the modification of general relativity
in the infrared region. In addition to above-mentioned cosmological motivations, 
SSB of general coordinate reparametrization invariance might lead to some resolution 
for cosmological constant problem \cite{Kirsch, 't Hooft}. Though the cosmological
constant problem involves quantum gravity, solving the problem seems to require
a low energy mechanism, for instance, a partial cancellation of vacuum energy
density stemming from quantum loops. In analogy with the Higgs mechanism for
conventional gauge theories, the Nambu-Goldstone mode mixes with the massless
graviton, thereby changing the vacuum structure of gravitational sector in a
non-trivial way.    

Recently, an idea of ghost condensation has put forward in order to obtain a
consistent infrared modification of general relativity \cite{Arkani}. In this scenario, the 
'$\it{unitary}$' propagating scalar field appears as the Nambu-Goldstone boson
for a spontaneously broken time-like diffeomorphism, and leads to the possibility
for resolving cosmological problems such as inflation, dark matter and dark energy.
The key point here is that a scalar ghost is converted to a normal scalar with
positive-energy excitations.  

More recently, 't Hooft has proposed an interesting Higgs mechanism for gravity
where massless gravitons '$\it{eat}$' four real scalar fields and consequently
become massive \cite{'t Hooft}. In this model, all the diffeomorphisms are broken 
by four real scalar fields spontaneously such that vacuum expectation values (VEV's) of
the scalar fields are chosen to the four space-time coordinates up to a
proportional constant by gauge-fixing diffeomorphisms. Of course, the number 
of dynamical degrees of freedom is unchanged before and after SSB. Actually, 
before SSB of diffeomorphisms there are massless gravitons of two dynamical 
degrees of freedom and four real scalar fields whereas after SSB we have massive 
gravitons of five dynamical degrees of freedom and one real scalar field 
so that the number of dynamical degrees of freedom is equal to six 
both before and after SSB as desired. Afterward, a topological term was
incorporated into 't Hooft model where an '$\it{alternative}$' metric tensor
is naturally derived and the topological meaning of the gauge conditions were
clarified \cite{Oda}. \footnote{Similar but different approaches have been
already taken into consideration in Ref. \cite{Oda1}.}  

The problem in the 't Hooft model is that a scalar field appearing after SSB is a 
non-unitary propagating field so that in order to avoid violation of unitarity
it must be removed from the physical Hilbert space in terms of some 
procedure. \footnote{More recently, models of gravitational Higgs mechanism 
without the non-unitary propagating scalar field were proposed
in Ref. \cite{Kaku2}.} 
A resolution for this problem was offered where one requires the energy-momentum 
tensor of the matter field couple to not the usual metric tensor but the modified 
metric one in such a way that the non-unitary scalar field does not couple to 
the energy-momentum tensor directly. 

Combining the ghost condensation scenario with gravitational Higgs mechanism,
one might be able to avoid emergence of the non-unitary scalar field in gravitational
Higgs mechanism since the annoying non-unitary scalar field has its roots in one 
time-like component (i.e., ghost) in four scalar fields, to which the ghost
condensation idea could be applied. Thus, it should be regarded that our
observation is an alternative method for removing the non-unitary propagating
scalar field in gravitational Higgs mechanism by 't Hooft.

In this short article, as the first step for proving this conjecture,
we study classical solutions in the unitary gauge in general ghost condensation
models. This analysis is needed for understanding in what background gravitational
Higgs mechanism arises. It turns out that depending on the form of scalar fields 
in an action, there are three kinds of classical, exact solutions, which are (anti-) 
de Sitter space-time, polynomially expanding universes and flat Minkowski space-time. 

We finally comment on gravitational Higgs mechanism in these models where we have massive 
gravitons of 5 degrees of freedom and 1 unitary scalar field (Nambu-Goldstone boson) 
after spontaneous symmetry breakdown of general coordinate reparametrization invariance. 
The models at hand are free from the problem associated with the non-unitary propagating 
mode. The detailed analysis will be reported in a separate publication.

\section{Original ghost condensation model}

{}For review and comparison with generalized models argued later, 
let us start with the original ghost condensation action in four space-time 
dimensions \cite{Arkani} \footnote{For simplicity, we have put the Newton constant 
$G$ at the front, so the dimension of ghost-like scalar field $\phi^0$ differs from that of
the original ghost condensation model, but it is easy to modify the dimension by
field redefinition.}:
\begin{eqnarray}
S = \frac{1}{16 \pi G} \int d^4 x \sqrt{-g} [ R - 2 \Lambda + f(X) ],
\label{Orig}
\end{eqnarray}
where $f(X)$ is a function of $X$ which is defined as
\begin{eqnarray}
X = g^{\mu\nu} \partial_\mu \phi^0 \partial_\nu \phi^0.
\label{X}
\end{eqnarray}

Recall that the ghost condensation scenario consists of three ansatzs:
The first ansatz amounts to the requirement that the function $f(X)$ has 
a minimum at some point $X_{min}$ such that
\begin{eqnarray}
f'(X_{min}) = 0, \  f''(X_{min}) > 0,
\label{Mini}
\end{eqnarray}
where $f'(X_{min})$, for instance, means the differentiation of $f(X)$ 
with respect to $X$ and then putting $X = X_{min}$. Note that the latter 
condition in (\ref{Mini}) ensures ghost condensation.

As the second ansatz, the ghost-like scalar field $\phi^0$ is expanded
around the background $m t$ as
\begin{eqnarray}
\phi^0 = m t + \pi,
\label{Pi}
\end{eqnarray}
where $\pi$ is the small fluctuation. This equation can be interpreted
as follows: Using a time-like diffeomorphism $\delta \phi^0 = \varepsilon^0$, 
one can take the '$\it{unitary}$' gauge $\pi =0$. 
In other words, $\pi$ is a Nambu-Goldstone boson associated 
with spontaneous symmetry breakdown of time translation. 
In this article, we consider only the case $\pi =0$ since we are interested in
classical solutions.

The last ansatz is natural from the second ansatz. That is, since 
time coordinate $t$ plays a distinct role from spacial coordinates
$x^i(i=1, 2, 3)$, it is plausible to make an assumption on background
space-time metric, which is of the Friedman-Robertson-Walker form
\begin{eqnarray}
ds^2 = - dt^2 + a(t)^2 d \Omega^2,
\label{FRW}
\end{eqnarray}
where $a(t)$ is the scale factor and $d \Omega^2$ is the spatial metric
for a maximally symmetric three-dimensional space.

The equations of motion are easily derived as follows:
\begin{eqnarray}
\partial_\mu (\sqrt{-g} g^{\mu\nu} f'(X) \partial_\nu \phi^0) &=& 0,
\nonumber\\
R_{\mu\nu} - \frac{1}{2} g_{\mu\nu} R + \Lambda g_{\mu\nu} &=& T_{\mu\nu},
\label{Eqs}
\end{eqnarray}
where the former equation is the $\phi^0$-equation of motion whereas
the latter equation is Einstein's equations where the stress-energy
tensor $T_{\mu\nu}$ is defined as 
\begin{eqnarray}
T_{\mu\nu} &\equiv& - \frac{16 \pi G}{\sqrt{-g}} \frac{\delta}{\delta g^{\mu\nu}}
[ \frac{1}{16 \pi G} \int d^4 x \sqrt{-g} f(X) ]        \nonumber\\
&=& - [ \partial_\mu \phi^0 \partial_\nu \phi^0 f'(X)
- \frac{1}{2} g_{\mu\nu} f(X) ].
\label{Stress}
\end{eqnarray}

The purpose of this paper is to look for classical solutions, so let us
solve these equations of motion under the ansatzs of ghost condensation.
First, note that under the second ansatz $\phi^0 = m t$ (recall that we put
$\pi = 0$ for classical analysis), the $\phi^0$-equation of motion requires
us that $f(X)$ should be minimized at $X_{min}$, that is,
\begin{eqnarray}
X_{min} \equiv X(\phi^0 = m t) = - m^2, \ f'(X_{min}) = 0.
\label{Ext}
\end{eqnarray}
Then, the stress-energy tensor $T_{\mu\nu}$ becomes proportional to
the metric tensor $g_{\mu\nu}$ and consequently Einstein's equations
reduce to the form where the cosmological constant is shifted by
this contribution 
\begin{eqnarray}
G_{\mu\nu} + \Lambda' g_{\mu\nu} = 0,
\label{Eins}
\end{eqnarray}
where we have defined the Einstein's tensor by $G_{\mu\nu} \equiv R_{\mu\nu} 
- \frac{1}{2} g_{\mu\nu} R$ and $\Lambda' \equiv \Lambda - \frac{1}{2} f(-m^2)$. 
Einstein's equations at hand are nothing but the conventional form of FRW cosmology, 
so the solutions are respectively de Sitter, flat Minkowski, and anti-de Sitter
universes according to the sign of the effective cosmological constant
$\Lambda'$.

\section{'t Hooft model with ghost condensation}

Recently, 't Hooft has proposed an interesting Higgs mechanism for gravity
where massless gravitons '$\it{eat}$' four real scalar fields and consequently
become massive \cite{'t Hooft}. In this model, all the diffeomorphisms are broken 
by four real scalar fields spontaneously. A key idea is that VEV's of the scalar fields 
are chosen to the four space-time coordinates by gauge-fixing diffeomorphisms. 
Therefore, three scalar fields are usual scalar ones while one scalar field has
a wrong-sign quadratic kinetic term, that is, it is a $\it{ghost}$ field
because of negative signature of the time coordinate. 
The problem in the 't Hooft model is that a scalar field appearing after SSB is a 
non-unitary propagating field, thereby leading to violation of unitarity.

In order to overcome this problem, it seems to be natural to apply the ghost condensation
idea to the 't Hooft model, which is our main motivation behind the present study. 
Our idea, whose detail will appear in a separate publication in future, is to
gauge-fix not all the diffeomorphisms but only spacial diffeomorphisms by
selecting VEV's of three scalar fields to be spacial coordinates and then use
the remaining time-like diffeomorphism to kill the non-unitary propagating mode.
However, even in our idea, classical configuration of four real scalar fields
must be taken to space-time coordinates up to an overall constant
\begin{eqnarray}
\phi^a = m x^\mu \delta_\mu^a,
\label{Gauge}
\end{eqnarray}
where the superscript $a$ runs over $0, 1, 2, 3$ as the space-time one $\mu$.
Thus, we shall take account of Eq. (\ref{Gauge}) as the classical configuration 
for four scalars. 

A new action of ghost condensation, which is inspired by gravitational Higgs mechanism, 
is of form
\begin{eqnarray}
S &=& \frac{1}{16 \pi G} \int d^4 x \sqrt{-g} [ R - 2 \Lambda + f(X) 
- g^{\mu\nu} \partial_\mu \phi^a \partial_\nu \phi^b \eta_{ab} ]  
\nonumber\\
&=& \frac{1}{16 \pi G} \int d^4 x \sqrt{-g} [ R - 2 \Lambda + F(X) 
- g^{\mu\nu} \partial_\mu \phi^i \partial_\nu \phi^j \delta_{ij} ] ,
\label{Main}
\end{eqnarray}
where $\eta_{ab}$ is the internal flat metric with diagonal elements $(-1, +1, +1, +1)$
and we have defined $F(X) \equiv X + f(X)$.
Let us note that the last term in the first equality is added to the original 
ghost condensation action (\ref{Orig}), or alternatively speaking from the side
of gravitational Higgs mechanism, the third term is introduced to the original action
of gravitational Higgs mechanism.

Let us seek classical solutions of this action. To do that, we first derive the
equations of motion. The $\phi^i(i = 1, 2, 3)$-, $\phi^0$-, and Einstein's equations
are respectively given by
\begin{eqnarray}
\partial_\mu (\sqrt{-g} g^{\mu\nu} \partial_\nu \phi^i) &=& 0,
\nonumber\\
\partial_\mu (\sqrt{-g} g^{\mu\nu} F'(X) \partial_\nu \phi^0) &=& 0,
\nonumber\\
G_{\mu\nu} + \Lambda g_{\mu\nu} &=& T_{\mu\nu},
\label{Eqs2}
\end{eqnarray}
where the stress-energy tensor now takes the form
\begin{eqnarray}
T_{\mu\nu} &=&  ( \partial_\mu \phi^a \partial_\nu \phi^b 
- \frac{1}{2} g_{\mu\nu} g^{\alpha\beta} \partial_\alpha \phi^a
\partial_\beta \phi^b ) \eta_{ab}  \nonumber\\
&-& ( \partial_\mu \phi^0 \partial_\nu \phi^0 f'(X)
- \frac{1}{2} g_{\mu\nu} f(X) ).
\label{Stress2}
\end{eqnarray}

Then, let us solve each equation in order. With the metric ansatz (\ref{FRW}) and
the assumption (\ref{Gauge}), the $\phi^i$-equation is trivially satisfied. 
Next, the $\phi^0$-equation reduces to the expression
\begin{eqnarray}
\partial_0 ( a(t)^3 F'(X) ) = 0,
\label{phi0}
\end{eqnarray}
whose validity requires us that one of ghost condensation ansatzs
should be automatically satisfied, i.e., $F'(X_{min}) =0$ since
$X_{min} \equiv X(\phi^0 = m t) = - m^2$ and $\partial_0 X_{min} = 0$.

Finally, Einstein's equations are cast to 
\begin{eqnarray}
3 (\frac{\dot a}{a})^2 - \tilde \Lambda &=& \frac{3 m^2}{2 a^2},
\nonumber\\
- 2 \frac{\ddot a}{a} - (\frac{\dot a}{a})^2 + \tilde \Lambda 
&=& - \frac{m^2}{2 a^2},
\label{Reduced Eins}
\end{eqnarray}
where we have defined $\tilde \Lambda = \Lambda - \frac{1}{2} F(-m^2)$.
Here we have made use of the expression of the stress-energy tensor 
which is obtained by inserting Eq. (\ref{Gauge}) to Eq. (\ref{Stress2})
\begin{eqnarray}
T_{\mu\nu} &=&  m^2 ( \eta_{\mu\nu} - \frac{1}{2} g_{\mu\nu} g^{ab} \eta_{ab} )
+ m^2 \delta_\mu^0 \delta_\nu^0 + \frac{1}{2} g_{\mu\nu} f(-m^2).
\label{Modified Stress2}
\end{eqnarray}
{}From the equations (\ref{Reduced Eins}), we can eliminate the terms involving $\dot a$
whose result is written as
\begin{eqnarray}
3 \frac{\ddot a}{a} = \tilde \Lambda.
\label{Reduced Eins2}
\end{eqnarray}
Depending on the value of $\tilde \Lambda$, the general solution for this equation
is classified into two types. When $\tilde \Lambda \neq 0$, we have one type of the
general solution
\begin{eqnarray}
a(t) = c_1 e^{\sqrt{\frac{\tilde \Lambda}{3}} t} 
+ c_2 e^{- \sqrt{\frac{\tilde \Lambda}{3}} t},
\label{Solution1}
\end{eqnarray}
where $c_1$ and $c_2$ are integration constants. However, since this solution
does not satisfy the first equation in Eq. (\ref{Reduced Eins}), it is
not in fact a solution. 

The other type of the general solution arises
when $\tilde \Lambda = 0$. Then, Eq. (\ref{Reduced Eins2}) reads 
$\ddot a = 0$, so we have the general solution
\begin{eqnarray}
a(t) = c (t - t_0),
\label{Solution2}
\end{eqnarray}
where $c$ and $t_0$ are integration constants. Substituting this solution into
the first equation in Eq. (\ref{Reduced Eins}), $c$ is fixed to 
$c = \pm \frac{m}{\sqrt{2}}$. As a result, the line element takes the form
\begin{eqnarray}
d s^2 = - d t^2 + \frac{m^2}{2} (t - t_0)^2 d \Omega^2,
\label{FRW2}
\end{eqnarray}
which describes the linearly expanding universe with zero acceleration.
Notice that this solution is a unique classical solution, so a flat
Minkowski space-time, for instance, is not a solution of this model.
Here it is worthwhile to mention that although the present universe seems to
be accelerating so that the linearly expanding universe is excluded from
WMAP experiment, the general solution (\ref{FRW2}) might be useful in describing 
the future or past status of universe.

\section{A general ghost condensation model}

In this section, we wish to generalize the model treated in the previous
section in the sense that the part including $\phi^i$ in the action 
(\ref{Main}) is not a linear function but a general function of 
$Y \equiv g^{\mu\nu} \partial_\mu \phi^i \partial_\nu \phi^j \delta_{ij}$. 
Through this generalization, it turns out that a flat Minkowski space-time 
also becomes a classical, exact solution of the equations of motion, 
which would be useful in applying this generalized model to QCD in future.

The action with which we start is given by
\begin{eqnarray}
S = \frac{1}{16 \pi G} \int d^4 x \sqrt{-g} [ R - 2 \Lambda + F(X) 
- H(Y) ],
\label{Generalized}
\end{eqnarray}
where '$\it{potential}$' term of $Y$ is now generalized to an arbitrary
function $H(Y)$. 
{}From this action, it is straightforward to derive the $\phi^i$-, $\phi^0$-, 
and Einstein's equations as before, which read
\begin{eqnarray}
\partial_\mu (\sqrt{-g} g^{\mu\nu} H'(Y) \partial_\nu \phi^i) &=& 0,
\nonumber\\
\partial_\mu (\sqrt{-g} g^{\mu\nu} F'(X) \partial_\nu \phi^0) &=& 0,
\nonumber\\
G_{\mu\nu} + \Lambda g_{\mu\nu} &=& T_{\mu\nu},
\label{Eqs3}
\end{eqnarray}
where the stress-energy tensor is of the form
\begin{eqnarray}
T_{\mu\nu} &=&  - \partial_\mu \phi^0 \partial_\nu \phi^0 F'(X) 
+ \partial_\mu \phi^i \partial_\nu \phi^j \delta_{ij} H'(Y) 
\nonumber\\
&+& \frac{1}{2} g_{\mu\nu} ( F(X) - H(Y) ).
\label{Stress3}
\end{eqnarray}

The line of argument for solving these equations of motion proceeds as
in the previous section. We find that the $\phi^i$-equation is trivially
satisfied. The $\phi^0$-equation again leads to the condition 
$F'(X_{min}) = 0$, which is equivalent to one of ansatzs for ghost condensation.
Then, Einstein's equations read
\begin{eqnarray}
3 (\frac{\dot a}{a})^2 - \tilde \Lambda &=& \frac{1}{2} H,
\nonumber\\
- 2 \frac{\ddot a}{a} - (\frac{\dot a}{a})^2 + \tilde \Lambda 
&=& \frac{m^2}{a^2} H' - \frac{1}{2} H.
\label{Reduced Eins3}
\end{eqnarray}

There are some exact solutions for these Einstein's equations by choosing
an appropriate form of $H(Y)$. We shall list up only two types of exact solutions,
which are polynomially expanding universes that are analogs of the linearly
expanding universe (\ref{Solution2}), and a flat Minkowski space-time.

Solutions describing the polynomially expanding universes are obtained 
by choosing $H(Y)$ to be
\begin{eqnarray}
H (Y) = K Y^n - 2 \tilde \Lambda,
\label{Pol-univ}
\end{eqnarray}
where $K$ is a constant. Then it is easy to show that the following
polynomially expanding universes are exact solutions for Einstein's equations
(\ref{Reduced Eins3}):
\begin{eqnarray}
a(t) = \sqrt{3} m ( \pm n \sqrt{\frac{K}{6}} (t - t_0) )^{\frac{1}{n}},
\label{Solution3}
\end{eqnarray}
which correspond to generalized solutions of the previous solution 
(\ref{Solution2}), for which $n = 1$ and $K = 1$.

It is remarkable that in this generalized ghost condensation model (\ref{Generalized}),
a flat Minkowski space-time becomes a classical solution as well whereas 
it is $\it{not}$ so in the 't Hooft model with ghost condensation
(\ref{Main}). Thus, this general model opens a new avenue to
gravitational Higgs mechanism, which will be utilized in QCD.
In fact, with the assumptions 
\begin{eqnarray}
\dot a(t) = \ddot a(t) = 0,
\label{Flat}
\end{eqnarray}
Einstein's equations (\ref{Reduced Eins3}) reduce to the form
\begin{eqnarray}
- \tilde \Lambda &=& \frac{1}{2} H,
\nonumber\\
\tilde \Lambda &=& \frac{m^2}{a^2} H' - \frac{1}{2} H.
\label{Reduced Eins4}
\end{eqnarray}
Thus, as a classical solution, we have a flat Minkowski
space-time if 
\begin{eqnarray}
H'(y) = 0,
\label{H}
\end{eqnarray}
where $y \equiv Y(\phi^i = m \delta_\mu^i x^\mu) = \frac{3 m^2}{a^2}$.
This equation holds true if we select, for instance, $H(Y)$ such that
\begin{eqnarray}
H(Y) = (Y - y)^N - 2 \tilde \Lambda,
\label{H2}
\end{eqnarray}
where $N$ must be more than unity.

\section{Discussion}

In this article, we have found classical solutions in the unitary gauge
$\pi = 0$ in general ghost condensation models, which are inspired by
gravitational Higgs mechanism by 't Hooft. It is shown that in the 't Hooft 
model with ghost condensation, there is a unique classical solution
with the form of the linearly expanding universe. On the other hand,
in a general ghost condensation model, we have a flat Minkowski space-time
in addition to the polynomially expanding universe's solutions. Thus,
this general model could lead to a candidate for gravitational Higgs
mechanism.

In order to show that the general model is really physically consistent, 
one has to prove that the non-unitary propagating scalar mode, which
appears in the original 't Hooft model for gravitational Higgs mechanism and 
is a serious problem, does not appear in the model at hand. Actually,
we have checked that there does not appear the non-unitary scalar in our
general model whose detail will be reported in a separate publication.

Although the detailed calculation is a bit involved, we can see why we have
a unitary theory by a simple argument. To do so, let us count the number of the 
dynamical degrees of freedom before and after SSB of diffeomorphisms. Before SSB, there are
massless gravitons of two dynamical degrees of freedom and four real
scalars whereas after SSB we have massive gravitons of five dynamical degrees 
of freedom and one scalar $\pi$, so the number of the dynamical degrees of freedom 
is the same before and after SSB of diffeomorphisms.
At this point, note that there is no room for the non-unitary propagating mode to exist in our 
general model. This is a reason why there is no non-unitary scalar field in this case.

However, it is still of importance to ask ourselves what symmetry is effective for
killing the non-unitary mode in our model. It is a time-like diffeomorphism
that kills the non-unitary mode. In this context, note that in the
background ansatz (\ref{Pi}) the time-like diffeomorphism is not broken
by transforming $\pi$ in a suitable manner. Thus, our general model
gives us not only a new model for ghost condensation but also
a nice model for gravitational Higgs mechanism.

\begin{flushleft}
{\bf Acknowledgement}
\end{flushleft}

We would like to thank K. Uryu for valuable discussions. Part of this work 
has been done when one of authors (I. O.) stayed at Dipartimento di Fisica, 
Universita degli Studi di Padova. We also wish to thank for a kind hospitality.

\vs 1   

\end{document}